# Architecting Network-Centric Software Systems: A Style-Based Beginning

Amine Chigani, James D. Arthur, and Shawn Bohner
*Department of Computer Science*
*Virginia Polytechnic Institute and State University*
*Blacksburg, VA 24061, USA*
*{achigani, arthur, sbohner}@vt.edu*

## Abstract

*With the advent of potent network technology, software development has evolved from traditional platform-centric construction to network-centric evolution. This change involves largely the way we reason about systems as evidenced in the introduction of Network-Centric Operations (NCO). Unfortunately, it has resulted in conflicting interpretations of how to map NCO concepts to the field of software architecture. In this paper, we capture the core concepts and goals of NCO, investigate the implications of these concepts and goals on software architecture, and identify the operational characteristics that distinguish network-centric software systems from other systems. More importantly, we use architectural design principles to propose an outline for a network-centric architectural style that helps in characterizing network-centric software systems and that provides a means by which their distinguishing operational characteristics can be realized.*

## 1. Introduction

The ubiquity of the network and the ability to deploy software over a network has given rise to network-centric software systems. As with many technology shifts, advancements in network technologies have changed the underlying assumptions for producing effective system solutions. The "system of systems" perspective now dominates much of the engineering as new systems are composed of multiple interconnected systems to support emerging missions. Moreover, the timeline of change has shifted from build custom (for optimum efficiency) to integrate and augment (for agile response to changing mission). The cycle time and economic implications are substantial as systems can be fielded in shorter periods of time integrating existing systems.

Software development efforts have seen an emergence of this new culture that focuses on the integration of existing and new software systems to tackle large and complex computing problems. A fundamental aspect of this culture is its substantial reliance on networked communications between the different elements of a system [14]. One reason behind this shift is the need to reach beyond tightly-coupled environments to access data and functionality that reside on remote systems that run on different platforms, and which are possibly owned and managed by different organizations. Another reason is the dynamic and complex structures of today's organizations, where the computing resources of an organization can span multiple national and international locations [7].

While NCO is a term coined by the Department of Defense [19], the concept has been applied in industry (Boeing, Lockheed-Martin, and others) with shaping systems for Agile operations, and in the National Aeronautics and Space Administration (NASA) with integrating systems for joint operations. We view the introduction of NCO as the pinnacle of the shift into the network-centric software development model. One reason is that many DOD contractors and research partners are now involved in contracts and research grants that focus on developing and providing tools and capabilities for all the supporting technologies of NCO.

Commercial organizations also recognize the necessity to streamline their business processes [25] through the integration of their diverse computing resources. Sharing the same strategic goals of NCO – which is the use of information technology to achieve business goals more efficiently – many commercial organizations have developed network-centric frameworks and architectures for their computing infrastructures.

Although network-centricity is widely recognized among software architects, a principled approach to understanding and mapping network-centricity concepts into the activities of architectural design is nascent. This has resulted in architectural solutions that do not comprehensively address the operational characteristics of software systems that are intended to be network-centric. In addition, there are several implications that network-centric software systems have on the field of software architecture. These implications must be addressed at the architectural level [5].

In this paper, we examine and discuss several aspects of network-centric software systems and address issues pertinent to their architectural design. To do so, we first examine the concepts of network centricity and describe how they have evolved to be recognized as fundamental concepts within the software architecture community. To emphasize its importance and uniqueness, we introduce a preliminary characterization of network-centric software systems and identify four distinguishing characteristics. We then present the issues relevant to the architectural design of such systems, e.g., security, scalability, and standardization. Finally, we employ architectural design principles to introduce and outline a new architectural style that help realize the operational characteristics of network-centric software systems.

We organize this report in the following manner. Section 2 includes an analysis of the concepts of network centricity. Section 3 discusses several challenges facing the field of software architecture. Section 4 details the reasons why these challenges must be addressed at the architectural level. Section 5 introduces our approach to addressing the problem of architecting network-centric software systems, which is through the construction of a new architectural style. Section 6 includes our conclusions. Finally, section 7 suggests future research work that needs to be done in order to complete the documentation of this new architectural style, and to establish its validity amongst other styles.

## 2. Concepts of Network Centricity

The term "network-centric" has many interpretations in the software engineering community. These different definitions underlie the chaotic approaches many organizations adopt to develop network-centric software systems. In this section, we present our understanding of network-centricity derived from the origin and background of the term "network-centric". The purpose is to construct a foundation upon which we can discuss the issue of architecting network-centric software systems.

### 2.1. Network-Centricity: Background

"Network-centric" is used loosely in many areas of the software engineering including software architecture [10, 17, 19]. Understanding the origin and background of this term enables us to use it more accurately and to describe what it means in the context of software architecture.

The term "network-centric" has gained a wide-spread use after the introduction of Department of Defense's network-centric operations. NCO is an emerging theory of war that seeks to translate an *information advantage* into a *competitive warfighting advantage* through the robust networking of well-informed, geographically-dispersed forces allowing new forms of warfighting organizational behavior [1]. NCO's basic tenets include:

- Utilizing technological advantages to support war fighters in the battlefield
- Networking all systems used by US armed forces
- Achieving shared awareness of the battlefield amongst all members of the US armed forces [24].

To achieve its goals, NCO depends on many technologies including network architectures, satellites, radio bandwidth, unmanned vehicles, nanotechnology, processing power, and, most importantly for this research, software.

Many argue that the military borrowed the concept of network-centricity from existing business models that software corporations, such as Oracle [20], have developed to integrate its diverse and distributed assets. Others argue that the concept originated from DOD and has found its way into industry as companies compete for government contracts to develop and provide tools, capabilities, and support mechanisms for NCO. Irrelevant to our discussion is whether the origin is NCO or industry; yet, it is critical that we understand the goals of network centricity within the context of warfighting, and its goals within the context of software-intensive systems.

Similar to NCO, network-centric software systems focus substantially on their communication element. To accomplish the ideals of NCO, these systems must accomplish effective application and data integration. This integration is achieved through taking various systems on different platforms (i.e., OSs), built with different object models, expressed using different programming languages [17], accessing different remote and local data repositories, and integrating them into robust systems for supporting critical business processes or scientific research programs.

A key reason for becoming network-centric is to be able to assemble software systems by integrating a mix of existing and new applications, and ensuring that the end-product is capable of integrating with other net-ready applications.

One of the more distinctive characteristics of network-centric software systems is that their communicating elements are, to a large extent, loosely-coupled sub-systems that work together to solve a larger and complex problem that cannot be solved by any individual sub-system. The idea of a network-centric software system is that of a "system of systems" [4] (Figure 1). Software engineers increasingly want and need to reach beyond tightly-coupled environments to access functionality on remote systems that are different in design, and which are perhaps owned and managed by other institutions [17].

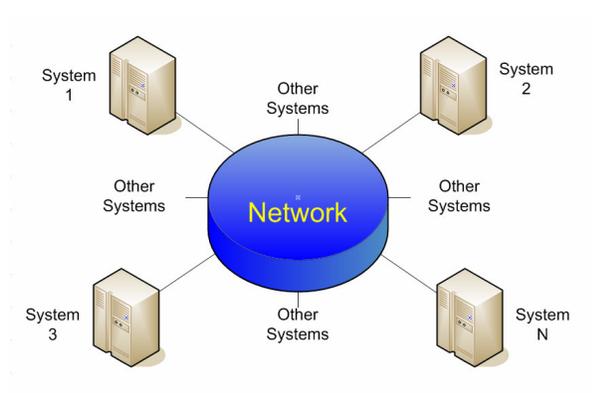

Figure 1. The notion of a "system of systems"

In the following section, we discuss in detail our understanding of the network-centric framework and its implications on the field of software architecture. This framework brings about challenges that affect software development, and in particular software architecture, and that require solutions at the architecture level.

## 2.2. A Framework for Software Development

Network-centric computing embodies the "information age invasion" [19] of many areas of science, art, business, education, government, and others. It represents a new way of thinking about how software engineers can produce software systems more effectively using network resources to efficiently integrate subsystems of information resources to respond to evolving mission requirements. This view is supported by the accumulating software architectures and frameworks that have network-centric characteristics. Balci *et al* survey existing software architectures and frameworks that claim to address the network-centricity issue, and that are adopted by a number of government and industry institutions [2].

Within the network-centric development framework, the focus is on two main constituents: 1) The network and communication types among the software system's components, and 2) the software system's behavior at runtime. Further, there are two important aspects in the network-centric development framework: A technology aspect and a human aspect. To elaborate, the network and communication types between the system's components correspond to the information technology side of the network-centric framework. Advances in networking technologies have been the drive that has led to the spread of a network-centric culture amongst software developers. On the other hand, the architects make the decisions on how elements of the system behave and communicate with one another in a networked setting to achieve a common objective. Therefore, the system behavior at runtime is driven by a human behavior manifested in the architects' design choices made during the software architecture design.

In our preliminary research, we have identified at least four characteristics that distinguish network-centric systems from other systems. A network-centric software system has:

- a "*system of systems*" perspective
- Has an underlying *networked configuration* that embodies the runtime environment on which the system's components interact and limits components' interaction to information exchange
- an *emergent*, *dynamic runtime behavior*, which means that the system's actual interacting components are not necessarily known until runtime and that the overall functionality of the system emerges from the collaborative behaviors of the components
- a *fluid, dynamically-defined decentralized control*, which means that control over the system's functionality is not necessarily owned by a particular component; rather, this control changes based on which function the system is performing and which component has initiated the system's execution. This control can be either strategic or tactical control

The network-centricity concepts have several implications on software engineering, and in particular, the software design (architecture) phase. In traditional software systems, a common theme has been that they are constructed as closed systems managed by single organizations [10]. Although some components are usually reused or obtained from other internal and/or external applications, the entire system comes under the control of the designer or architect once integrated. For such systems, architectural design has often resulted in architectures that do not easily allow any dynamic behavior at runtime [10]. In this setting, architects must know before hand where components will be located and how to interface with them. Architects have control over all components and therefore are able to predict their behavior at runtime.

However, this assumption is void within the network-centric model. A network-centric software system may have a central objective but not a centralized control. A leading application of the concepts of network-centricity is represented in the Internet [17]. The Internet is a collaborative network of networks that exhibits an emergent behavior that is a result of its complex architecture. Nevertheless, the Internet structure is facilitated by a minimal set of standards [10] in the form of protocols that describe how to exchange data over the network. These protocols are independent of the hardware or software applications that use the Internet. More importantly, adherence to these protocols is voluntary with no central authority that posses coercive power. Websites, web services, and other Internet-based activities are managed by their individual organizations and the decision to join and/or leave the Internet network is solely in the hands of that organization.

Architects of such systems face an emerging set of challenges. Their implications go beyond creating a single reference architecture that will support the design and implementation of many kinds of network-centric software systems. Rather, these challenges induce the need for a general approach to designing software architectures for these kinds of systems. We anticipate that this approach must be in terms of a new architectural style, added to the reservoir of existing styles, which will enable architects to design systems that answer to both the demands of network centricity and their respective problem domain. The reason is that the concept of a style defines the common features of a family of software architectures for a particular class of systems [8], network-centric software systems comprise one such class.

# 3. Related Issues to Network-Centricity

Many challenges that the software architecture community faces are not specific to network centricity. This is because the field of software architecture itself is still maturing. The following are the most dominant ones that we have identified by investigating software systems and software architectures that exhibit network-centric characteristics.

## 3.1. Standardization

The first challenge is the related need to develop software architectures that flexibly accommodate applications and services provided by various developers. An emerging trend in software development efforts is that systems are composed out of a mix of local and remote computing capabilities, requiring architectural support that accommodates interoperability, modifiability, connectivity, security, and other desirable operational qualities [8]. Thus, we argue that this support should come in the form of an architectural style that facilitates the generation of systems using a dynamically-formed coalition of distributed resources. More specifically, new standards (similar to Internet protocols) need to be established for building new components and making existing one net-ready.

## 3.2. Scalability

Building on the analogy between network-centric software systems and the Internet, a second challenge emerges: The need for scalable architectures that can evolve and that can handle component complexity and variability similar to architecture of the Internet. Network-centric software systems, being systems of systems, incorporate different components that require different architectural representations and various forms of communication. While many of the existing architectural styles will likely apply, the details of their application will need to change. Thus, we see that there is a need to define a new architectural style that accommodates these changes.

For instance, implicit invocation is a widely-accepted method of designing software systems. Implicit invocation is a style of software architecture in which a system is organized around event handling – broadcasting and subscribing to events. On one hand, this style allows for heterogeneous components to be integrated into systems that have low-coupling and high-cohesion with are two indispensable qualities of any software system. On the other hand, architects must make assumptions about certain qualities which are crucial to the system such as the reliability of event delivery and routing of messages. In a network-centric model, all such assumptions are uncertain [10]. Therefore, we are further convinced that we need novel techniques that allow architects to design these systems in such a way that it accommodates their dynamic growth.

### 3.3. On-Demand Composition

The third challenge is the need to develop architectures that permit end users to form their own system composition. With the rapid growth of the Internet, an increasing number of users are in a position to assemble and tailor services. Such users may have minimal technical expertise, and yet, will still want a sufficiently strong guarantee that the parts will work together in the ways they expect [10].

Architects must to find ways to support such needs for network-centric software systems. A network-centric architecture has to encompass characteristics that facilitate the generation of systems which are modifiable and that support an on-demand integration of new components.

### 3.4. Robust Connectivity

The fourth challenge that faces designers of network-centric software systems is the need for a robust infrastructure, which supports computing through a large number of independent, heterogeneous, distributed, dynamically-integrated components. For instance, the Internet infrastructure supports a broad range of resources such as primary information, communication mechanisms, web applications, services, and many others [10]. A common characteristic among these resources is independence – both operational and managerial. They can join and leave the network at will, they can invoke other resources and can be invoked, and, most importantly, they evolve independently of each other. Similarly, a network-centric software system must have an underlying infrastructure that facilitates a decentralized control over the system elements. Elements are selected and composed based on the task that needs to be carried out.

Due to the intrinsic complexity of automating the selection and composition process, architects must focus on the interface requirements between the elements of a network-centric software system. Within a network-centric model, architects do not necessarily have implementation knowledge about the components that are developed by other entities. In addition, the integration of incorporated components may be unfeasible if these components have static interface specifications. For instance, the integration of a component packaged to interact via remote procedure calls with a component packaged to interact via shared data can be a difficult task [10].

These are added challenges for architectural design. Creating an architectural style that facilitates the consideration of these challenges at the architecture level seems a reasonable proposition.

### 3.5. Security

Security models focus on the secure exchange of information among components of a system to meet the requirements defined by the problem domain [26]. In the case of network-centric software systems, the intense reliance on networked communications brings about more security risks and concerns. Security cannot be an added feature to the system; it needs to be built into the system [15]. Therefore, architectures that support the generation of network-centric software systems need to provide the capability to have security technologies built into the appropriate elements of the system infrastructure.

### 3.6. Test and Evaluation

The concern over test and evaluation issues is nearly as old as the concept of NCO [22]. In their book on Network-Centric Operations [1], Alberts, Garstka, and Stein discuss the implications of the concept stating that: "Testing systems will become far more complex since the focus will not be on the performance of individual systems, but on the performance of federations of systems." This leads to the conclusion that traditional engineering techniques for evaluating network-centric software architectures will not be able to completely meet the network-centric software systems test and evaluation need. Traditional techniques are likely necessary, but by no means sufficient.

We do not claim the list of challenges outlined above is comprehensive. As the software architecture community gains more insight into network-centric software systems, we believe more challenges will be identified.

## 4. Constructing Network-Centric Software Systems

### 4.1. Where Do We Start?

While the term "network-centric software systems" has no unified formal definition, the phenomenon is widespread and generally recognized. In this paper, we have identified several characteristics of these systems. Network-centric software systems represent an emergent class of software systems that are built from components that are large scale systems in their own right. The main difference from monolithic large scale systems is the independence of software components. This difference results in a greater emphasis on interface design and communication standards than in conventional system architecting and engineering. A question that raises itself is *"at what phase of development must this difference be addressed?"*

Since the components of a network-centric software system are often developed independently, the system emerges only through the interaction of the components. The architect must express an overall structure (architecture) largely through the specification of communication standards. Thus, to accurately address the development challenges that network-centric software systems lead to, additional architectural techniques must be employed.

## 4.2. Importance of Architectural Design

The literature is rich with definitions of software architecture [3, 7, 8, 10, 21]. They all agree on the fact that software architecture is a set of components and connectors among them. Figure 2 depicts a typical representation of a legacy software architecture of a system [3]. We easily notice that the figure does not completely convey what we need to know about the system. From an architecture perspective, many things cannot be inferred from the diagram:

- The nature of the elements: Objects, tasks, processes, distributed components, etcetera.
- Their responsibilities and function within the system
- The meaning of the connectors: Communication, control, data-sharing, synchronization, invocation, use, or some combination of these or other relations
- The importance of the layout. Are they all on the same line because there is no room, or for some other reason?

Questions such as the ones we have raised must be addressed unless we know precisely what each of the elements in the diagram stands for. This diagram does not show a software architecture in any way that could be useful to the user. However, we can thoughtfully say that it is a start.

Architectural design has long been accepted as an essential ingredient of a well designed software system. It includes the first attempt to model the high level structure of a system. A study conducted by Barry Boehm empirically confirms that investments in architecting are increasingly necessary for large scale projects such network-centric software systems [5].

Within architectural design, architects think about software in different ways: Software modules, components and the connectors among them, or allocation of software components to their environment. Clements *et al* refer to these perspectives as viewtypes [8]. Relevant to our discussion here is the component-and-connector viewtype, C&C viewtype for short. This viewtype corresponds to the way architects look at the software system as a set of elements and their interactions at runtime. The C&C viewtype is a relevant way of describing and documenting an architectural style.

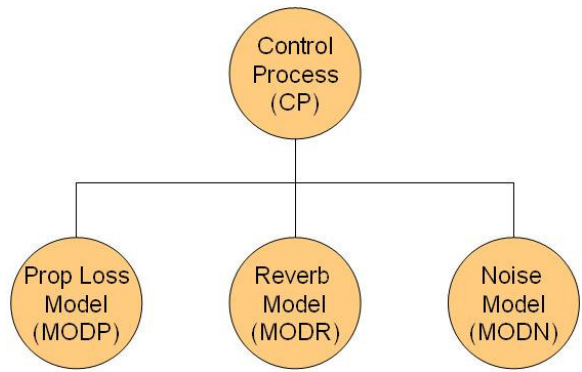

Figure 2.   Typical representation of a software architecture [3]

## 4.3. Example Viewtype for Architectural Style

To better understand what the C&C viewtype provides for expressing an architectural style, we present an abbreviated example of a service-oriented architectural style. This indicates the type of style information that would be conveyed using the C&C viewtype.

| Service-Oriented Architectural Style Guide | | |
|---|---|---|
| *Elements* | *Component:* | Service component |
| | *Type:* | Service |
| | *Component:* | UDDI Registry |
| | *Type:* | Registry |
| | *Component:* | SOAP message |
| | *Type:* | Remote procedure call |
| | *Component:* | Invocation |
| | *Type:* | Invoke procedure |
| | *Connector:* | Functional component |
| | *Type:* | Active Process |
| | *Connector:* | User Interface |

| | Type: | GUI Process |
|---|---|---|
| *Relations* | *Attachment* relation associates service interfaces with SOAP message connector. Service interface information is encapsulated as service description in WSDL. *Attachment* relation associates the registry with SOAP messages to access or publish service information. | |
| *Computational Model* | Services are registered in the registry. Service invokers acquire registered service information and produce corresponding adapters to invoke services. | |
| *Properties of the elements* | Service: <br>• *Name:* should suggest its functionality <br>• *Type:* defines service type provided as a passive unit that responds to caller <br>• *Interface Properties:* depend on the type of the network communication <br>• *Attachment:* not persistent. <br>UDDI Registry: … <br>SOAP Message: … <br>Remote Procedure Call: … <br>Functional Component: … <br>User Interface Component: … | |
| *Topology* | Each service has at least one adapter. Services can be invoked concurrently or in synchronously depending on adapter organization. | |

Table 1. Summary of the service-oriented style

## 4.4. The Case for a Network-Centric Architectural Style

Architectural styles emerge as formal architectural approaches after architects have been using these styles for a while [21]. Once a style proves to be effective in solving a particular design problem, architects then formalize its definition and documentation, and make it available as a choice in the architectural design space. Our research in the area of network-centric software systems has led us to recognize a trend in the way network-centric software architectures/systems are applied to a certain class of problems. These situations usually entail large, distributed systems, configured in a system of systems for missions on a short timeline, integrated via advanced communications to provide the relevant interoperability. While multiple styles combined can describe these configurations of systems, this trend is generally different from existing trends/styles. Therefore, we propose the recognition and formalization of this new architectural style. We argue that existing styles, individually, cannot respond to the emerging challenges of network-centric software systems; that is application and data (both new and legacy) integration. Thus, we have begun defining the elements of this new style by taking advantage of the beneficial characteristics of many existing styles to describe the diverse nature of network-centric software systems.

## 5. A Network-Centric Architectural Style

An essential part of documenting a new style is to develop a style guide that records the specialization and constraints the style imposes on its elements and their interactions. This section presents an outline of the *network-centric* architectural style. The format of documentation is based on the C&C viewtype structure used in [8]. Figure 3 depicts an interaction between two elements of a software system by means of a networked communication which the essential characteristic of network-centric software systems.

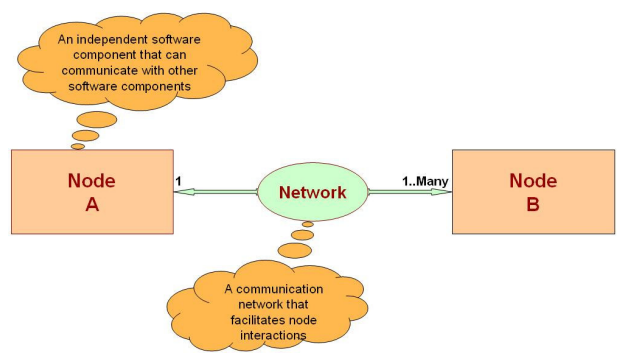

Figure 3. Depiction of a basic tenet of network-centric software systems - networked interaction

### 5.1. Style Elements

The core element of network-centric software systems is their network component. Therefore, we catalog the network-centric style using the C&C viewtype. The elements of the network-centric style are composed of components and connectors. Components are peer systems, client applications, servers, concurrent units, and any software component that can be viewed as an independent system regardless of size or functionality. We call these components *nodes*. Connectors are request/reply, invokes-procedure, data exchange, message passing, control, and other types of communication that allow components to cooperate within a network in solving a larger task than what each of them can individually handle. We call these kinds of connectors: *links*.

Nodes of such systems are independent sub-systems that collaborate to solve a common business or scientific problem. For instance, an online store requires many collaborating entities to carry out an online transaction. It requires a secure, usable, reliable web interface that allows customers to shop for products. It also needs a credit card processing capability to handle the information of the customer in a secure mode. And finally, it needs a shipping and handling component that takes care of placing the items in the hands of the customers. This is a simplified version of a business transaction that requires at least three main nodes. These nodes collaborate to carry out the business transaction. It is not necessary that the three components are built and managed by the same organization; however, it is necessary that, at runtime, they work together in various ways to perform their required task that assists in solving the problem at hand. The choice of how they interact is left to the designer, and is based on the type of the components (servers, peers, etc), whether the components belong to the same institution, or whether they are services obtained from one or more providers. The components work concurrently and interact with one another in various ways, both symmetric and asymmetric. There are no constraints on how these components interact from this perspective.

Links focus mainly on the network communication type. Different domains use different types of network communications. This often depends on whether software components communicate with hardware components other than those that are hosting them. An example of this kind of hardware is a data sensor.

## 5.2. Style Relations

In this style, there is only one kind of relationship between elements: Attachment. Nodes communicate with one another based on their relationships and the underlying network used for such communication. The main characteristic of this particular attachment is that it is dynamic. The network-centric style draws on distributed communicating nodes that interact with one another dynamically to carry out a global task. However, the style can be applied recursively. It can be applied to represent how distributed system components interact, and can also represent individual nodes that are themselves network-centric in nature.

## 5.3. Style Computational Model

Network-centric software systems are composed of nodes and links. Nodes are connected dynamically based on the current need to carry out a task. Nodes communicate with one another via links. These links represents different communication mechanisms.

We characterize nodes as abstractions so that they can represent components from other existing architectural styles as well as new kinds of components that might be recognized in the future. Similarly, links are abstractions of the communication mechanism and, as such, support both existing and evolving networking technologies, e.g., IPv6.

## 5.4. Summary

Clearly, we have not discussed the network-centric style in a complete detail; we have, however, provided a foundation for doing so. This style consists mainly of nodes and links that facilitate communications among the nodes. The analogy of an application of network style is that of a dynamic data structure. When we define a dynamic data structure such as linked list for instance, we do not know what an instance of that data structure will look like at runtime. However, when we design this data structure, we can define how new nodes can be added and removed. Similarly, the network-centric style allows architects to identify the overall structure of how nodes work together, and what kind of links can be used to interact with one another independent of the participating nodes and links at runtime. The network-centric style also helps in moving critical design decisions to the level of architecture, which in turns makes it possible to address certain quality attributes and to perform risk analyses to avoid misallocation of resources. Table 1 provides a summary of the proposed network-centric style.

| Network-Centric Architectural Style Guide | |
|---|---|
| *Elements* | • *Component types:* Independent software nodes with dynamic interfaces<br><br>• *Connector types:* Links which facilitates communication between nodes |
| *Relations* | *Attachment* relation associates a node interface with another node interface |

| Computational Model | Nodes are connected by means of links between them. A node can communicate with multiple nodes simultaneously |
|---|---|
| Properties of the elements | Node: <br><br> • *Name:* should suggest its functionality <br><br> • *Type:* defines general functionality of the element <br><br> • *Interface Properties:* depend on the type of the network communication <br><br> Link: <br><br> • *Name:* should suggest the nature of the interaction <br><br> • *Type:* defines the nature if the interaction and the required parameters <br><br> • *Other Properties:* depending on network communication type, it may include protocol of interaction and performance values |
| Topology | There is no topological model to this style. Nodes connects and disconnects to other nodes at will |

Table 2. Summary of the network-centric style

## 6. Conclusions

Network-centric software systems embody an answer to the pressing need to integrate existing software assets with newly developed applications, and to the growth in size of software-intensive systems that are being developed. They also exemplify a new way of thinking about software systems. Network-centric software systems are the outcome of an inevitable shift from developing a system of statically-distributed resources to a system of dynamically-distributed components that are owned and managed by different entities, and that provide task specific services [11] that can be used to achieve a larger goal. This shift has brought about a need for new architectural approaches to design such systems. In this paper, we have discussed the characteristics of network-centric software systems based on the concepts of network-centric operations. Further, we have detailed the importance of the field of software architecture and the architectural activities that can help in deploying quality software systems. Using the mature standards of software architecture, we have formulated and propose a new architectural style that describes an emerging class of systems – network-centric software systems.

## 7. Future Work

The network-centric architectural style guide is by no means complete. It is part of our ongoing research activity. Currently, we are finalizing the style guide for this new architectural style. A number of activities remain to be completed to substantiate our proposed style. Amongst these activities is to add examples and scenarios describing when and how to use this style. Also, much work needs to be done in terms of comparing the network-centric style with existing ones to validate its standalone status. Distinguishing this style from existing ones will help architects better understand and more easily adopt this style in their practices. Currently, we are looking at one non-defense industry (Intelligent Transport Systems) that we intend to use as an example to further demonstrate the usefulness of the concepts of network-centricity.